\documentclass[a4paper]{epl}
\usepackage{graphicx}
\usepackage{amssymb}

\title{Long transients and cluster size in globally coupled maps}
\author{Guillermo Abramson\thanks{E-mail:
\email{abramson@cab.cnea.gov.ar}}} \institute{Consejo Nacional de
Investigaciones Cient{\'\i}ficas y T{\'e}cnicas, Centro At{\'o}mico Bariloche,
8400 S.C. de Bariloche, Argentina } \pacs{05.45.-a}{Nonlinear
dynamics and nonlinear dynamical systems.}
\pacs{05.45.Xt}{Synchronization; coupled oscillators.}

\begin{document}

\maketitle

\begin{abstract}
We analyze the asymptotic states in the partially ordered phase
of a system of globally coupled logistic maps. We confirm that,
regardless of initial conditions, these states consist of a few
clusters, and they properly belong in the ordered phase of these
systems. The transient times necessary to reach the asymptotic
states can be very long, especially very near the transition line
separating the ordered and the coherent phases. We find that,
where two clusters form, the distribution of their sizes
corresponds to windows of regular or narrow-band chaotic behavior
in the bifurcation diagram of a system of two degrees of freedom
that describes the motion of two clusters, where the size of one
cluster acts as a bifurcation parameter.
\end{abstract}

Systems formed by globally coupled logistic maps
\cite{kaneko89,kaneko90,chate92} are a paradigm of the rich
phenomenology arising when opposing tendencies compete: the
nonlinear dynamics of the maps, which in the chaotic regime tends
to separate the orbits of different elements, and the coupling,
that tends to synchronize them. The synchronization of dynamical
elements from an undifferentiated state makes these systems
particularly interesting in many interdisciplinary applications,
such as network of optical elements, networks of Josephson
junctions, ecological systems and social
behavior~\cite{perez92,dominguez93,heagy94}.

The equations describing the evolution of a system with $N$ globally
coupled elements are:
\begin{equation}
x_i(t+1) = (1-\epsilon)f[x_i(t)]+\frac{\epsilon}{N}\sum_j f[x_j(t)]
\end{equation}
where $f[x]=r\, x (1-x)$ is the standard logistic map, and
$\epsilon$ is a global coupling parameter. Different phases
characterize the behavior of the system in the space spanned by
$r$ and $\epsilon$. The usual nomenclature for these phases is
based in the one introduced by Kaneko \cite{kaneko90}. For large
$r$ and small $\epsilon$ the behavior is chaotic in space and time
(``turbulent''). For $\epsilon$ large enough there is a
synchronized state with all the elements evolving in the same
trajectory (``coherent'' phase). In intermediate regions a number
of phases can be found where the systems collapses in clusters of
synchronized motion. Each cluster behaves as a single element, an
effective reduction of the number of degrees of freedom taking
place. Two kinds of these ``ordered'' phases have been
recognized, according to the structure of the basins of
attraction of the system. If almost all initial conditions are
attracted to states with a few clusters, the phase is
``ordered.'' If attractors with a few clusters coexists with
many-cluster attractors, the phase is termed ``partially
ordered'' or ``glassy.'' These have been observed in the
transition regions between the turbulent and the ordered phase
(type I), and between the ordered phase and the coherent phase
(type II). We have studied this last region (typically at $r=4$
and $\epsilon=0.3~\mbox{to}~0.5$) observing that the asymptotic
state of the system consists of a few clusters (typically only
two) regardless of initial conditions. The ``partially ordered''
phase comes out to be a transient phenomenon.

\begin{figure}
\centerline{\resizebox{10cm}{!}{\includegraphics[bb=15 15 819
582,clip]{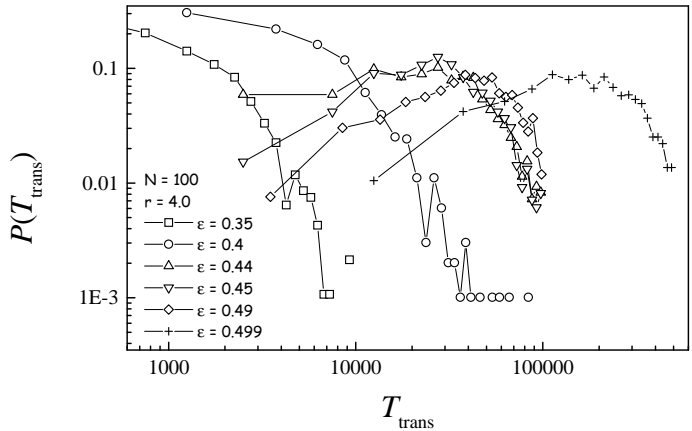}}}
\caption{Distribution of transient time for systems with
different values of the coupling parameter.}
\label{ttrans}
\end{figure}

The transient time needed to reach this asymptotic state can be
extremely long, and this may be the reason why this region of
parameter space had been incorrectly classified as glassy. This
has been observed by Manrubia and Mikhailov in a recent
paper~\cite{manrubia99}. Some years ago, also Xie and Cerdeira
\cite{xie96} observed that there is a transition from the one
cluster coherent state to a two clusters state at $r=4$,
$\epsilon=0.5$, without stressing that this was in contrast to
the \emph{glassy} phase believed to exist in this region of
parameter space. We have confirmed both these observations. In
fig.~\ref{ttrans} we show the distribution of the time needed to
reach a state of two clusters, calculated from several thousand
initial conditions, in a system with 100 elements. When the
coupling approaches the value $\epsilon=0.5$, the distribution
moves to longer times, and a maximum appears. At $\epsilon>0.5$
the coherent state is reached in a few time steps.

\begin{figure}
\centerline{\resizebox{10cm}{!}{\includegraphics[bb=15 15 819
582,clip]{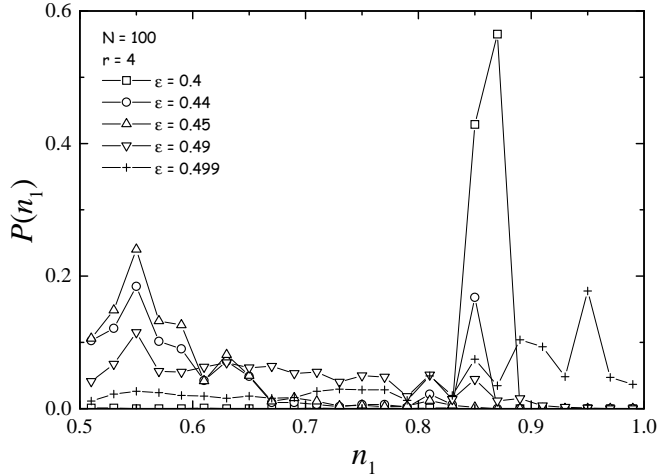}}}
\caption{Distribution of cluster size for systems with different
values of the coupling parameter.}
\label{tamanio}
\end{figure}

The two clusters that form asymptotically are not always of the
same size. Different initial conditions are attracted to unequal
clusters. The distribution of the size of one of the clusters is
shown in fig.~\ref{tamanio}. We plot $P(n_1)$, where $n_1=N_1/N$
is the size of the biggest of the two clusters relative to the
system size $N$. Five curves are shown with different symbols,
corresponding to different values of the coupling, ranging from
$\epsilon=0.4$ to $\epsilon\lesssim 0.5$. It can be seen that the
size distribution depends strongly on the value of the coupling.
The case with $\epsilon=0.35$ is not shown in the figure for
clarity, but it is worthwhile to observe that in such a case most
of the states have equally sized clusters, with $n_1\approx 0.5$;
there is also a smaller peak at $n_1\approx 0.85$. For
$\epsilon=0.4$, the peak at $n_1\approx 0.5$ is drastically
reduced and most of the states contain a cluster of size
$n_1\approx 0.85$. At still grater values of $\epsilon$ this peak
also disappears. Meanwhile, a well defined peak appears in the
distribution at $n_1 \approx 0.55$. This, in turn, disappears for
$\epsilon\approx 0.5$.

\begin{figure}
\centerline{\resizebox{10cm}{!}{\includegraphics{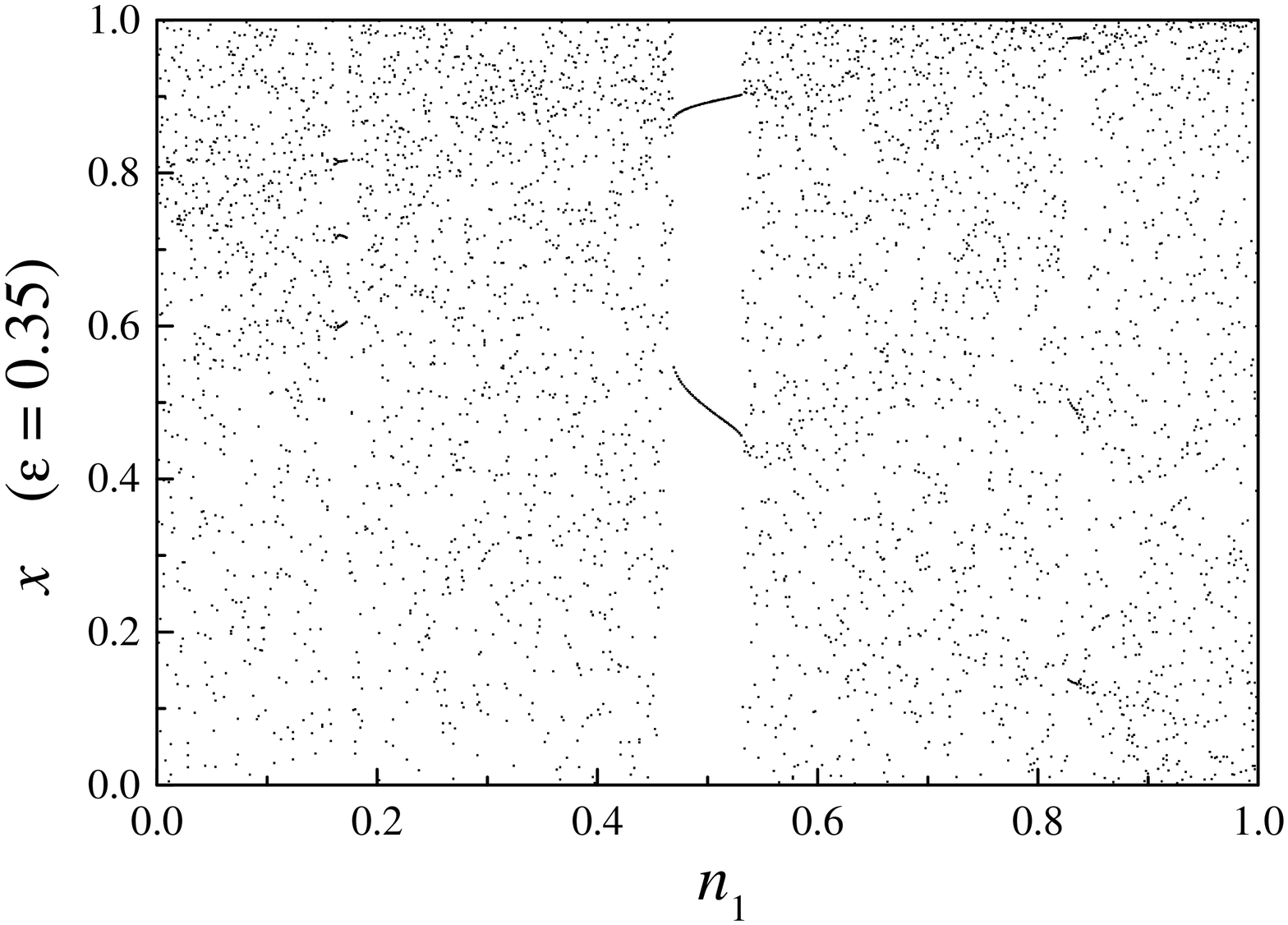}}}
\caption{Bifurcation diagram of one element in a system of two
elements [eq. (\protect\ref{dosclus})] representing two clusters
of the complete system. The cluster size acts as bifurcation
parameter. The other parameters are $\epsilon=0.35, r=4.0$.}
\label{035low}
\end{figure}

We can also analyze the type of trajectory followed by each
cluster in the asymptotic state. For example, at $\epsilon=0.35$
the two clusters are of the same size, as mentioned before. Each
one of them is almost always engaged into a period-2 orbit.
Occasionally, one of the clusters is found in a chaotic regime
with three narrow bands (similar to a period-3). These phenomena
turns our attention to the system of two degrees of freedom
corresponding to the asymptotic state of the full system. After
collapsing in two clusters, the system can be described as (see
\cite{xie96}, eq. (7)):
\begin{eqnarray}
x_1(t+1)&=&(1-\epsilon)f(x_1)+\epsilon\, n_1 f(x_1)+\epsilon
(1-n_1) f(x_2) \nonumber \\
x_2(t+1)&=&(1-\epsilon)f(x_2)+\epsilon\, n_1 f(x_1)+\epsilon
(1-n_1) f(x_2),
\label{dosclus}
\end{eqnarray}
where each coordinate describes the motion of one of the clusters.
For fixed values of  $\epsilon$ and $r$, we can consider the
cluster size $n_1$ as a bifurcation parameter of the
system~(\ref{dosclus}). In fig.~\ref{035low} we show the
bifurcation diagram of this system, $x_1$ \emph{vs} $n_1$, for
$\epsilon=0.35, r=4.0$.

\begin{figure}
\centerline{\resizebox{10cm}{!}{\includegraphics{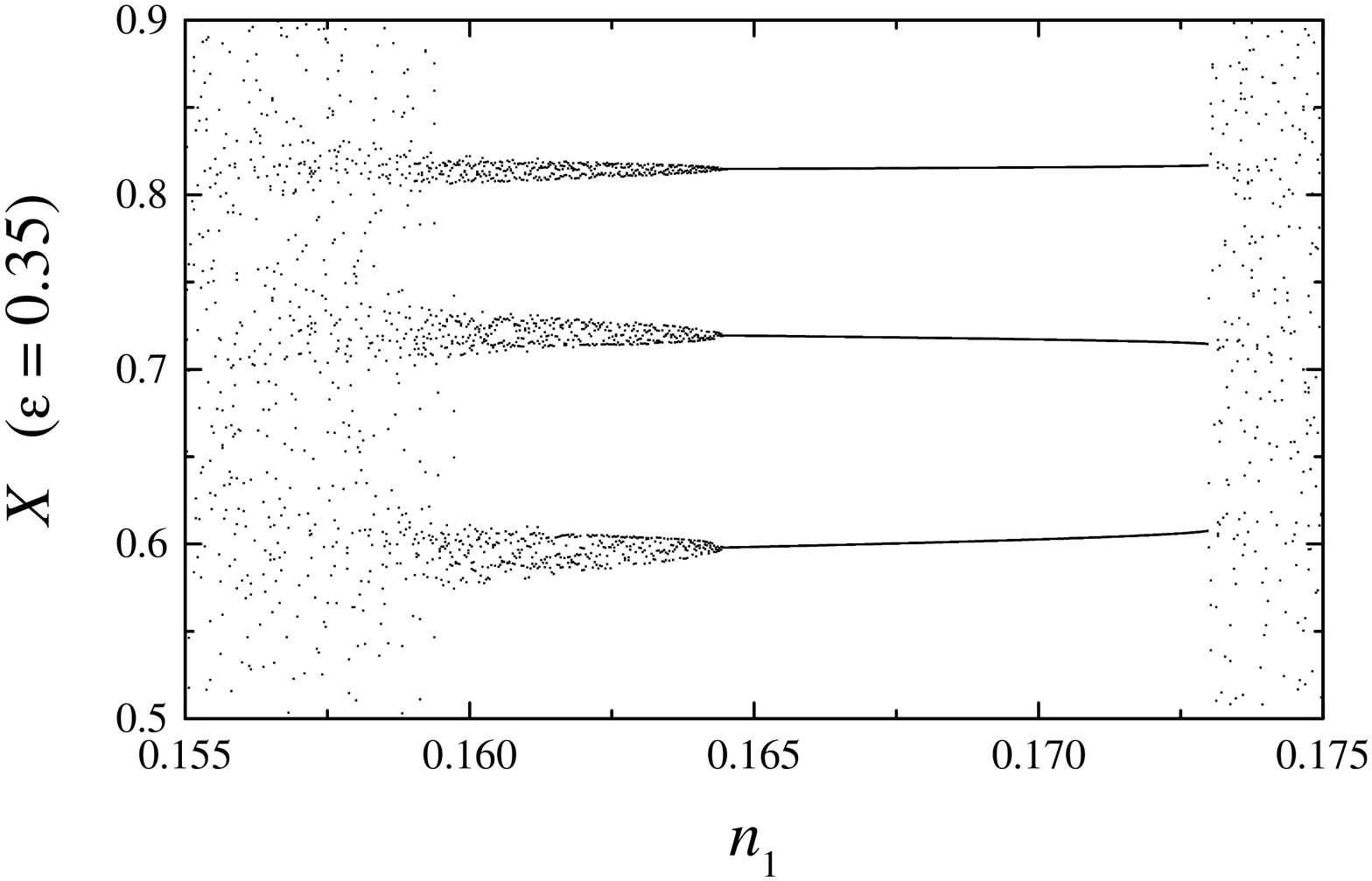}}}
\caption{A detailed view of fig~\protect\ref{035low}, showing the
second widest window, with a period-3 and three chaotic bands.}
\label{p3}
\end{figure}

A wide period-2 window is apparent in fig.~\ref{035low} around
$n_1=0.5$. Two narrower windows can be found at cluster sizes
$n_1\approx 0.82$ and $n_1\approx 0.18$. In these, the existence
of three bands can be observed---see fig.~\ref{p3} for an
expanded view. Another, very narrow window (it cannot be seen in
the figure at the current scale) is present near $n_1=0.65$. It
is a period-15, effectively observed in a few realizations of the
full system. For a finite system only some of these states are
accessible, since $N_1$ must be an integer.

\begin{figure}
\centerline{\resizebox{10cm}{!}{\includegraphics{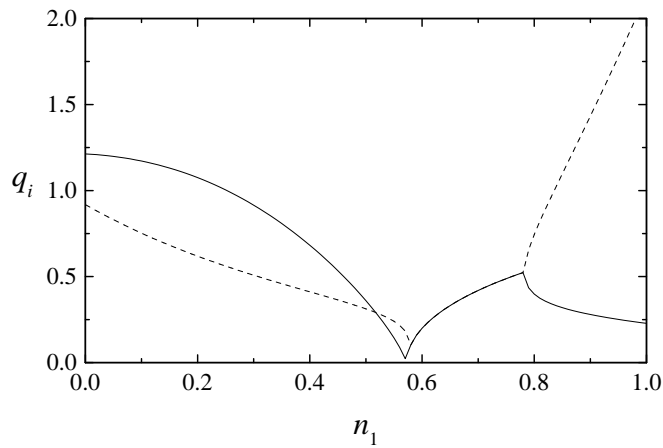}}}
\caption{Absolute value of the eigenvalues of the transformation
of the two degrees of freedom system. The eigenvalues are
evaluated on the period-2 asymptotic states attained by a
realization of the full system ($N=1000$, $r=4$ and
$\epsilon=0.35$) that collapses in two clusters with $n_1=0.53$.}
\label{eigenvalues}
\end{figure}

For different values of $\epsilon$ the structure of the
bifurcation diagram is different. The windows of regular behavior
are progressively narrower at increasing coupling. At
$\epsilon=0.5$ no more windows of regular motion are found, the
ordered phase finishes, and the coherent phase starts. We have
found that the two-clusters state always corresponds to the
windows of more regular behavior. Wide windows are represented by
high peaks in the distribution of cluster size, and narrow windows
by small peaks. The precise relationship between both magnitudes
is very difficult to establish. It is clear, of course, that the
mere existence of these windows in the two-degrees of freedom
system does not guarantee that the corresponding states are an
attractor of the system \emph{before} it collapses in a two
clusters state. Nevertheless, knowing the final states in some
particular cases we can analyze some properties of the collapsed
system. For example, when the final state of the two clusters is
a period-2 of states $x_1$ and $x_2$, the map is described by the
transformation:
\begin{eqnarray}
G_1(x_1,x_2)&=&F_1[F_1(x_1,x_2),F_2(x_1,x_2)] \nonumber \\
G_2(x_1,x_2)&=&F_2[F_1(x_1,x_2),F_2(x_1,x_2)]
\label{per2}
\end{eqnarray}
where
\begin{eqnarray}
F_1(x,y)&=&[1-\epsilon(1-n_1)]f(x)+\epsilon(1-n_1)f(y)\\
F_2(x,y)&=&\epsilon\, n_1 f(x)+(1-\epsilon n_1)f(y).
\end{eqnarray}
Now we can analyze the eigenvalues $q_i$ of the Jacobian matrix:
\begin{equation}
J=\left(
\begin{array}{cc}
\frac{\partial G_1}{\partial x_1} & \frac{\partial G_1}{\partial x_2}
\\
\frac{\partial G_2}{\partial x_1} & \frac{\partial G_2}{\partial x_2}
\end{array}\right)
\end{equation}
as functions of the cluster size $n_1$. A typical case is shown in
fig.~\ref{eigenvalues}. Here we have used a system with 1000
elements, at $r=4$ and $\epsilon=0.35$, and waited until it
collapses into a two-cluster attractor. Its orbit was then used
in the analysis of the Jacobian matrix. It fig.~\ref{eigenvalues}
we can see a range of cluster sizes where the modulus of both
eigenvalues is lower than one, as required by the stability
criterion. Within this range, both eigenvalues attain a minimum
near the cluster size corresponding to the actual partition of
the full system ($n_1=0.53$). The strength of the stability seems
to favor the actually realized partition, since the asymptotic
states can be found near the most stable ones. It has to be
stressed that the period-2 could---in principle---be found as a
fixed point of the map (\ref{per2}), from which the cluster size
could be predicted. However, this approach has proven to be very
difficult to treat analytically.

To summarize, the numerical analysis presented here confirms that
the partially ordered phase of type II of globally coupled
logistic maps has very long transients. After these transients, an
asymptotical state of only two clusters is typically attained. The
system of two clusters can be analyzed by means of a bifurcation
diagram where the cluster size acts as a bifurcation parameter.
The bifurcations suggest that the asymptotic cluster distribution
corresponds to the windows of regular or narrow-band motion of the
system of two clusters. This is backed up by the fact that the
eigenvalues of the Jacobian of the collapsed system, analyzed on
numerically realized orbits, show a minimum near the
corresponding cluster size.

\acknowledgments The author acknowledges fruitful discussions
with D. H. Zanette, and thanks H. A. Cerdeira for comments on the
manuscript.

\end{document}